\documentclass{article}

\usepackage[dblblindworkshop, final]{neurips_2025}

\workshoptitle{Deep Learning For Code in the Agentic Era}

\usepackage[utf8]{inputenc} 
\usepackage[T1]{fontenc}    
\usepackage{hyperref}       
\usepackage{url}            
\usepackage{booktabs}       
\usepackage{amsfonts}       
\usepackage{nicefrac}       
\usepackage{microtype}      
\usepackage{xcolor}         
\usepackage{minted}

\usepackage{tikz}
\usetikzlibrary{positioning, arrows.meta, shapes, fit, calc}
\usepackage{adjustbox}
\usepackage{subcaption}

\title{Agint: Agentic Graph Compilation for Software Engineering Agents}

\author{%
  Abhi Chivukula \\
  Agint\\
  \texttt{Abhi@AgintAI.com} \\
  \And
  Jay Somasundaram \\
  Agint\\
  \texttt{Jay@AgintAI.com} \\
  \And
  Vijay Somasundaram \\
  Agint\\
  \texttt{Vijay@AgintAI.com} \\
}

\begin{document}

\maketitle

\begin{abstract}
  LLM-based coding agents are increasingly common, but still face challenges in context management, latency, reliability, reproducibility, and scalability. We present \textbf{Agint}, an agentic graph compiler, interpreter, and runtime that incrementally and hierarchically converts natural-language instructions into typed, \textit{effect-aware} code DAGs (Directed Acyclic Graphs). Agint introduces explicit \textit{type floors} (TEXT $\to$ TYPED $\to$ SPEC $\to$ CODE) grounded in semantic graph transformations and a hybrid LLM/function JIT runtime, enabling dynamic graph refinement, reproducible and optimizable execution, speculative evaluation, and interoperability with existing developer tools. Agint’s typed graph bindings ensure reliability and enable \textit{concurrent composition of codebases by construction}, supporting accelerated development using smaller, faster models for lower latency, more efficient context use, and higher throughput. Hierarchical compilation allows scalable graph edits, while the graph structure provides reproducibility and efficient parallel generation. Agint provides a composable Unix-style toolchain: \textbf{dagify} \textit{(DAG compiler)}, \textbf{dagent} \textit{(hybrid JIT runtime)}, \textbf{schemagin} \textit{(schema generator)}, and \textbf{datagin} \textit{(data transformer)} for realtime, low-latency code and dataflow creation. Human developers and coding agents refine graphs through the \textit{Agint CLI}, while non-technical users leverage \textit{Agint Flow GUI} with visual editing, conversational refinement, and debugging to promote prototype agentic workflows to production code. This continuous co-creation model lets teams prototype quickly, refine seamlessly, and deploy reliably, bridging natural language, compiler methods, and developer tooling to enable a new generation of composable, team-centric coding agents at scale.
\end{abstract}

\section{Introduction}

Large Language Models (LLMs) have enabled coding agents that translate natural language into executable code~\citep{chen2021evaluating,li2022alphacode}. However, in real-world software engineering, there are still issues to address: syntactic errors, hallucinated functionality, and training-distribution quirks often require significant human correction. Models struggle with project-specific references in limited context, degrade with very long contexts, and either run slowly (large models) or prove unreliable (small models), especially for domain-specific tasks and structured outputs~\citep{gu2025aiforse,key2022trustworthy}.

These issues intensify in multi-agent and multi-developer settings. Concurrent edits introduce contention and cascading errors, as agents lack reliable coordination~\citep{qian2024chatdev,hong2023metagpt}. Unstructured generation is fast but fragile, while structured generation is more controllable but slower, particularly for specialized models without a fast inference provider~\citep{muendler2025typeconstrained}. Multi-model delegation through standards such as Multi-Client Proxy (MCP) adds token overhead, latency, variability, and dependency on opaque third-party implementations.

Software engineering rarely involves code alone: it requires coordination with datasets, APIs, and external connectors. Current agents cannot robustly integrate across these resources while staying efficient~\citep{liu2024agents4se,wang2024agents}. Even when producing step-by-step plans, execution is sequential and error-prone, with limited ability to parallelize or adapt to runtime values. Long-reasoning models exacerbate this by delaying early parallelism. Auxiliary tasks such as browsing, querying, or data transformation interrupt workflows and create dependencies that are difficult to track. In practice, code generation is inseparable from \textit{data organization} and \textit{workflow orchestration}~\citep{li2024autokaggle}.

While traditional programming elevated data structures and algorithms as the foundation of computation, the agentic era demands \textit{semantic data organization} and \textit{agentic workflow orchestration}. \textbf{Agint} addresses this by compiling natural-language intent into typed, effect-aware Directed Acyclic Graphs (DAGs) that unify code generation, data handling, workflow orchestration, and agent coordination. By grounding interactions in graphs, Agint ensures concurrency safety and composability across code, data, and tools, turning code generation from a long-running for-loop of fragile one-shot predictions into a structured process of agentic co-creation.

The graph form enables \textit{recursive concurrent task decomposition}: complex problems are partitioned into independent subgraphs that can be solved in parallel, refined incrementally, and recomposed without global bottlenecks~\citep{besta2024graph,yao2023treeofthought}. Because Agint treats workflow design as a compilation problem, heavy models can be used for \textit{ahead-of-time agentic compilation}, laying out the workflow structure, while lightweight, low-latency models execute and refine individual nodes~\citep{zenkner2025taskdecomposition}.

After the workflow skeleton is created, it undergoes two concurrent processes: \textit{refinement} and \textit{resolution}. Refinement enriches each node with unstructured contextual information—examples, constraints, documentation, or domain knowledge—that guides downstream generation without altering the type system~\citep{le2024codechain}. Resolution, by contrast, upgrades the node types in a structured manner: a natural-language node is successively promoted into a function signature, then into a formal specification, and finally into an executable code implementation~\citep{chen2024funcoder}. Importantly, Agint does not require waiting until full resolution to run a workflow: nodes can be executed in simulated or mock modes at intermediate stages, enabling early testing, speculative execution, and validation before compilation. Together, refinement and resolution preserve the graph's structure while progressively attaching typed bindings and contextual detail, making execution modular, reproducible, and tightly integrated across the workflow.

Agint's DAGs localize changes to graph nodes, allowing small, fast models to solve subproblems in parallel, escalating to larger models only when needed~\citep{arora2024masai,zhang2024dei}. Each node inherits context from its ancestry, minimizing global overhead. Classical compiler and runtime techniques—JIT-style optimization, speculative graph evaluation—further improve efficiency and reliability. This reduces context overhead, increases throughput, accuracy, and enables transparent, reproducible workflows~\citep{hu2025qualityflow}.

Agint thus bridges flow-based programming with agentic autonomy, providing a foundation for scalable, collaborative coding agents. In this paper, we introduce Agint’s compiler, interpreter, and runtime, and show how they enable reliable, scalable, and user-friendly workflows for code generation and data transformation.


\section{System Overview}

Agint implements a multi-stage graph compilation pipeline that transforms natural language specifications into executable code through a hierarchy of typed intermediate representations. Unlike traditional code generation approaches that operate on isolated functions, our system compiles entire workflows as directed acyclic graphs (DAGs), where nodes progress through six distinct type floors: \texttt{TEXT} $\to$ \texttt{TYPED} $\to$ \texttt{SPEC} $\to$ \texttt{STUB} $\to$ \texttt{SHIM} $\to$ \texttt{PURE}.

Each type floor represents a computational abstraction with well-defined semantics:
\begin{itemize}
\item \textbf{TEXT}: Natural language descriptions with implicit data flow, and task dispatch if the underlying execution LLM supports it
\item \textbf{TYPED}: Nodes acquire explicit type signatures (\texttt{PrimitiveType} system constraining to \texttt{str}, \texttt{int}, \texttt{float}, \texttt{bool} and their lists)
\item \textbf{SPEC}: Formal specifications with pre/post conditions and behavioral constraints
\item \textbf{STUB}: Function signatures with stub implementations
\item \textbf{SHIM}: Hybrid execution nodes containing both deterministic code and AI-synthesized virtual functions
\item \textbf{PURE}: Fully resolved executable code with no AI dependencies
\end{itemize}

The key innovation lies in the executability of intermediate representations. A \texttt{TYPED} node can execute through prompt chaining, while a \texttt{SHIM} node employs hybrid execution—deterministic code paths invoke AI-generated virtual functions (\texttt{VIRTUALSTUB}, \texttt{VIRTUALSHIM}, \texttt{VIRTUALPURE}) on demand. This enables incremental development where partially-specified workflows remain runnable throughout the compilation process.

\subsection{Compilation Architecture}

The compilation pipeline employs three core mechanisms:

\textbf{1. Type-Directed Resolution}: Each node maintains a \texttt{RESOLUTION\_STATE} (\texttt{UNRESOLVED}, \texttt{IN\_PROGRESS}, \texttt{PARTIALLY\_RESOLVED}, \texttt{FULLY\_RESOLVED}) that guides the compiler's strategy. Resolution preserves semantic equivalence while upgrading representations—a \texttt{TYPED} node's behavioral specification becomes explicit pre/post conditions in \texttt{SPEC}, which translate to function signatures in \texttt{STUB}.

\textbf{2. Locality-Preserving Transformation}: The DAG structure localizes compilation context to node neighborhoods. When resolving a node from \texttt{SPEC} to \texttt{STUB}, the compiler considers only immediate dependencies and dependents, not the entire graph. This enables parallel compilation where independent subgraphs resolve concurrently without global synchronization.

\textbf{3. Fallback Synthesis}: When direct compilation fails (e.g., ambiguous specifications, complex domain logic), the system employs three fallback strategies:
\begin{itemize}
\item \textit{Decomposition}: Complex nodes split into multiple simpler nodes
\item \textit{Virtual Functions}: Unresolvable logic becomes \texttt{VIRTUALSHIM} nodes for runtime synthesis
\item \textit{Deferred Compilation}: Nodes marked for resolution in subsequent passes
\end{itemize}

\subsection{Runtime Execution Model}

The runtime implements a hybrid execution strategy that seamlessly blends compiled code with AI-powered synthesis. Three execution modes support different performance/flexibility tradeoffs:

\textbf{Prefine Mode}: Runtime optimizes the nodes' function before execution while it's waiting for input from currently running nodes. This mode optimizes runtime codegen quality when code is later dynamically generated.

\textbf{Dynamic Mode}: Nodes execute with just-in-time synthesis. When encountering a \texttt{VIRTUALSHIM}, the runtime generates implementations using upstream execution context. This enables adaptive behavior where node implementations specialize based on actual data flow.

\textbf{Predict Mode}: Speculative execution runs multiple node implementations in parallel. The runtime predicts likely execution paths and pre-generates likely function input arguments and pre-emptively executes the function, hiding synthesis, and execution latency through prediction.

\subsection{System Components}

The architecture comprises specialized subsystems interconnected through well-defined interfaces:

\textbf{Flyte} provides unified LLM orchestration with hierarchical structured generation. It implements prompt registries, multi-provider routing, and automatic failover. The integration with Hydantic inference enables decomposition of complex outputs into parallel subtasks, reducing generation latency by 3-10× for large structured outputs with multiple independent fields.

\textbf{Dagify} serves as the graph compiler, implementing the type floor progression through a series of specialized resolvers. Each resolver (\texttt{PlainToTypedResolver}, \texttt{TypedToSpecResolver}, etc.) handles one transition in the type hierarchy. The compilation process supports incremental refinement where nodes can be individually upgraded without full recompilation.

\textbf{Dagent} implements the hybrid runtime with effect-aware execution. It tracks all side effects (filesystem, network, database operations) through an effect monad, enabling safe rollback and reproducible execution despite non-deterministic AI components. The runtime partitions DAGs into independent subgraphs that execute concurrently on separate workers. Additionally, Dagent is capable of runtime tool use of Dagify, Schemagin, Datagin, and MCP.

\textbf{Schemagin/Datagin}: Schemagin generates typed schemas from natural language, while Datagin handles data transformation and synthesis. Both integrate through Agilink (unified data connectivity) and Agitransfer (versioned artifact management), providing a complete prompt-to-database pipeline that mirrors the code compilation process.

\subsection{High-Level Architecture and Data Flow}

\begin{figure}[t]
  \centering
  \begin{adjustbox}{max width=\linewidth}

  \begin{tikzpicture}[
    node distance=0.7cm and 1.2cm,
    box/.style={rectangle, draw, rounded corners, thick, align=center, minimum width=2.4cm, minimum height=0.8cm, fill=blue!5},
    gbox/.style={draw, rounded corners, inner sep=0.4cm, thick},
    line/.style={dotted, -Latex, shorten >=2pt, shorten <=2pt, thick}
    ]

    \node[box, anchor=north] (schemagin) at (-6.4,0) {schemagin};
    \node[box, below=of schemagin] (datagin) {datagin};
    \node[box, below=of datagin] (pagint) {pagint};

    \node[box, anchor=north] (hydantic) at (0,0) {hydantic};
    \node[box, below=of hydantic] (flyte) {flyte};
    \node[box, below=of flyte] (flyteAPI) {flyteAPIService};
    \node[box, below=of flyteAPI] (agintAPI) {agintAPIService};

    \node[box, anchor=north] (dagify) at (6.4,0) {dagify};
    \node[box, below=of dagify] (dagent) {dagent};

    \node[gbox, fit=(schemagin)(datagin)(pagint),
    label={[yshift=0.2cm]above:\textbf{Semantic Data Organization}}] (g_semdata) {};

    \node[gbox, fit=(hydantic)(flyte)(flyteAPI)(agintAPI),
    label={[yshift=0.2cm]above:\textbf{Augmented Inference}}] (g_infer) {};

    \node[gbox, fit=(dagify)(dagent),
    label={[yshift=0.2cm]above:\textbf{Agentic Workflow Orchestration}}] (g_workflow) {};

    \node[box, below=2.9cm of agintAPI] (agitransfer) {agitransfer}; 
    \node[box, left=3.2cm of agitransfer] (agilink) {agilink};       
    \node[box, right=3.2cm of agitransfer] (agiwrite) {agiwrite};    
    \node[box, below=0.6cm of agilink] (agicat) {agicat};
    \node[box, below=0.6cm of agiwrite] (agit) {agit};

    \coordinate (leftAlign)  at (g_semdata.west |- agitransfer.center);
    \coordinate (rightAlign) at (g_workflow.east |- agitransfer.center);

    \node[gbox, fit=(agilink)(agitransfer)(agiwrite)(agicat)(agit)(leftAlign)(rightAlign),
    label={[yshift=0.35cm]above:\textbf{AI Utilities}}] (g_utils) {};

    \draw[line] (schemagin) -- (datagin);

    \draw[line] (schemagin.east) -- ($(flyte.west)+(0,0.18)$);
    \draw[line] (datagin.east) -- ($(flyte.west)-(0,0.18)$);
    \draw[line] (pagint.east) to[bend left=8] (flyte.west);

    \draw[line] (dagent) -- (dagify);
    \draw[line] (dagify.west) -- (flyte.east);
    \draw[line] (dagent.west) to[bend left=8] (flyte.east);

    \draw[line] (schemagin.south east) .. controls +(0.9,-0.9) and +(0.0,1.0) .. (agilink.north);
    \draw[line] (datagin.south east)  .. controls +(1.1,-1.1) and +(0.2,1.0) .. (agilink.north);

    \draw[line] (agicat.north) -- (agilink.south);
    \draw[line] (agiwrite.north west) to[out=180, in=20] (agilink.north east);

    \draw[line] (schemagin.south east) .. controls +(1.2,-1.2) and +(-1.3,1.2) .. (agitransfer.north west);
    \draw[line] (datagin.south)        .. controls +(0,-1.0)  and +(-1.0,1.0)  .. (agitransfer.west);
    \draw[line] (pagint.south)         .. controls +(0,-1.0)  and +(-1.0,1.0)  .. (agitransfer.west);
    \draw[line] (dagify.south)         .. controls +(0,-1.2)  and +(0.8,1.0)   .. (agitransfer.north east);
    \draw[line] (dagent.south)         .. controls +(0,-1.2)  and +(0.4,1.0)   .. (agitransfer.north);

    \draw[line] (flyte.north) -- (hydantic.south);
    \draw[line] (flyteAPI.north) -- (flyte.south);

  \end{tikzpicture}
  \end{adjustbox}
  \caption{Architecture Overview}
\end{figure}
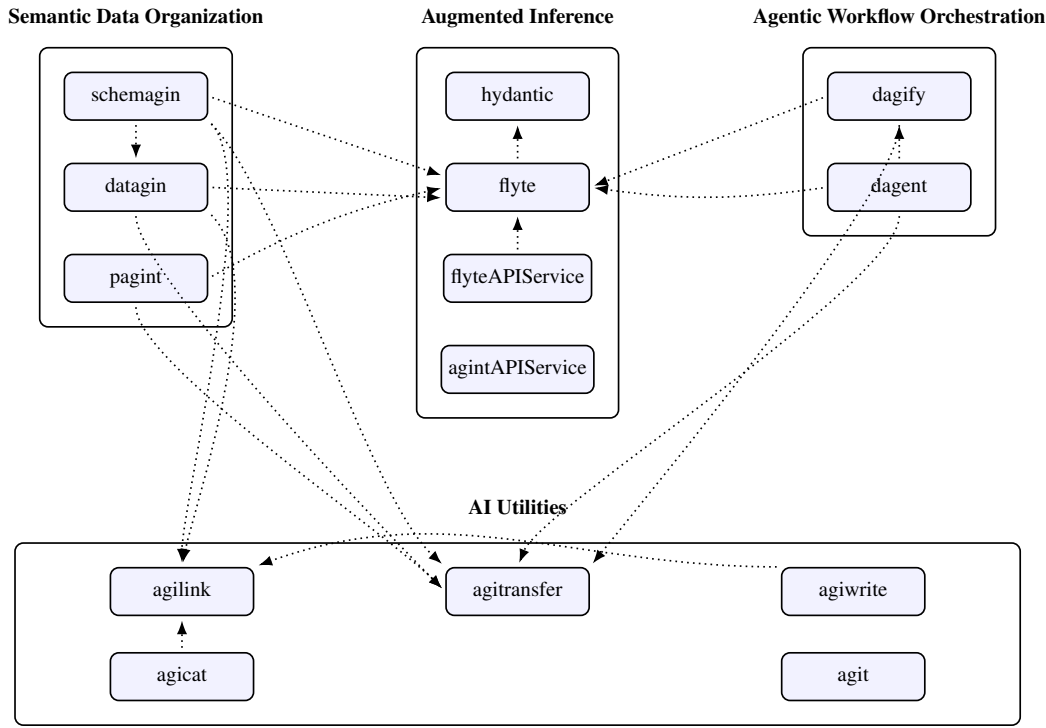

\noindent\textbf{Data Flow}: \emph{Dagify} generates and evolves DAGs through \emph{Flyte} and produces versioned YAML/JSON artifacts. \emph{Dagent} executes DAGs (or interprets them on the fly), invoking LLMs only through Flyte, and interacting with \emph{Agilink} for remote database connectivity and \emph{Agitransfer} for version control. \emph{Schemagin} primarily writes schemas (SQL/JSON/YAML/DBML) that \emph{Agilink} can read; \emph{Datagin} \emph{consumes schemas from Schemagin} (when provided) and performs \emph{ingest/synthesis/transform}, reading/writing data through \emph{Agilink}. All components share consistent CLI patterns and Pydantic-like models registered with Flyte which are dynamically converted to BAML.

\subsection{How the System Works}

\subsubsection{Dagify - Compose/Refine/Resolve/Compile}
Users or agents provide an NL specification; Dagify composes a \emph{Plain} DAG, refines nodes with context, resolves them to \emph{Typed} and \emph{Spec} nodes, and optionally compiles to \emph{Code} DAGs or target-specific builds (e.g., CrewAI, LangChain, WDL, BAML). All LLM calls (multi-shot, with fallback) are mediated by Flyte, with structured outputs validated against registered models.

\subsubsection{Dagent - Execute/Interpret}
Dagent executes a precompiled DAG, interprets and runs dynamically, or compiles-then-runs. The runtime supports speculative execution and effect-aware transactions/rollback, and partitions independent subgraphs for parallelism.

\subsubsection{Schemagin \& Datagin}
Schemagin generates schemas and visualizations; Datagin ingests unstructured inputs, synthesizes test data, or transforms datasets against provided schemas. Both route I/O through Agilink (DBs) and Agitransfer (files) as needed.

\subsection{Flyte: Key Benefit and Why It Is Fast}

\textbf{Centralized, async, multi-provider structured generation.} Flyte is the single LLM gateway: it manages prompt/structured-output registries, routes across providers, and executes \emph{asynchronously by default}, enabling high concurrency under rate limits. Crucially, Flyte integrates \textbf{Hydantic} (below) so complex structured outputs are generated via \emph{hierarchical, parallel subcalls} with focused context, which improves latency and reduces required context with useful accuracy. Together, multi-provider routing + async orchestration + Hydantic’s parallel structured filling are why Flyte-backed operations are robust and fast in practice.

\subsection{Hydantic in Context (What It Does)}

\textbf{Hydantic}, extends Pydantic with \emph{hierarchical decomposition} of nested models into levels (simple fields, Pydantic objects, nested Hydantic models). Flyte uses this to \emph{parallelize} field- or sub-model generation with \emph{focused prompts} per branch, decreasing per-call context and improving fill accuracy. The result is \emph{concurrent, progressive} population of complex outputs (often yielding substantial speedups (relative to a more expensive model) for models with many independent fields). Hydantic is automatically detected by Flyte’s frameworks and used when a Hydantic model is registered as the structured output. (The name Hydantic is a portmanteau of Huygens + Pydantic, because the inference frontier propagates through the Pydantic model via Huygens principle.)

\subsection{What Schemagin and Datagin Actually Do (and Their Data Flows)}

\textbf{Schemagin} converts NL descriptions into database schemas, refines existing schemas, and renders multiple formats (SQL, JSON, YAML, DBML) with visualization support (GraphViz/D2/ASCII). Its outputs are \emph{schema artifacts} that \emph{Agicat} and \emph{Agiwrite} can read and write, which \emph{Datagin} can consume as \emph{schema input}.

\textbf{Datagin} supports three roles: \emph{ingest} (extract structure from un/semistructured sources), \emph{synthesis} (generate schema-conformant test data), and \emph{transform} (convert between formats/schemas). Datagin \emph{optionally} takes Schemagin’s schema as input and performs reads/writes through \emph{Agilink} and \emph{Agiwrite} and spawns fully formed databases. Both Schemagin and Datagin register prompts/structured outputs with \emph{Flyte} and adhere to the shared CLI/async generation patterns.


\section{Agint CLI}

Agint implements a Unix-style toolchain where each CLI tool performs a specific function in the compilation and execution pipeline. The tools communicate through structured text formats (YAML/JSON/DBML) and a unified addressing system (\texttt{agilink://}), enabling flexible composition while maintaining type safety and reproducibility.

\subsection{Workflow Orchestration Tools}

\textbf{dagify} serves as the DAG compiler, transforming natural language specifications into progressively refined workflow graphs. It implements four primary operations:
\begin{itemize}
\item \texttt{compose}: Generates initial DAGs from natural language at a specified type floor
\item \texttt{refine}: Adds contextual information to nodes without changing types
\item \texttt{resolve}: Upgrades DAGs to higher type floors through AI-assisted compilation
\item \texttt{compile}: Produces executable artifacts for target platforms (WDL, CrewAI, LangChain, BAML)
\end{itemize}

\textbf{dagent} provides the runtime environment for DAG execution. It supports multiple execution strategies:
\begin{itemize}
\item \texttt{validate}: Performs static analysis and type checking
\item \texttt{optimize}: Tunes prompts and parameters against test datasets
\item \texttt{execute}: Runs DAGs with configurable JIT modes (prefine, dynamic, predict)
\item \texttt{interpret}: Generates and executes workflows directly from natural language
\item \texttt{synthesize}: Combines plan generation, compilation, and execution in one pass
\end{itemize}

Execution is constrained by explicit tool whitelists (\texttt{--tools}), ensuring reproducibility and preventing unintended side effects.

\subsection{Data Management Tools}

\textbf{schemagin} generates and manipulates database schemas from natural language. It provides schema authoring (\texttt{compose}), iterative refinement (\texttt{refine}), and multi-format visualization (\texttt{visualize}) supporting SQL, JSON, YAML, and DBML outputs with ASCII, GraphViz, and D2 rendering.

\textbf{datagin} handles data transformation workflows through three core operations:
\begin{itemize}
\item \texttt{ingest}: Extracts structured data from unstructured sources
\item \texttt{synthesize}: Generates realistic test data conforming to schemas
\item \texttt{transform}: Converts data between schemas with rule generation
\end{itemize}

\textbf{agicat} and \textbf{agiwrite} provide bidirectional synchronization between schema definitions and live databases, enabling round-trip engineering between design and deployment.

\textbf{pagint} offers row-level data operations, generating individual records or augmenting existing datasets with missing fields. It complements datagin's batch operations with fine-grained, internet-aware data synthesis.

\subsection{Integration Architecture}

The \texttt{agilink://} URI system provides a typed addressing layer that abstracts over storage backends. Components reference schemas, datasets, and transformations uniformly regardless of whether they reside in files, databases, or memory. This abstraction enables:

\begin{itemize}
\item Type-safe data flow between tools
\item Transparent caching and versioning
\item Distributed execution without code changes
\item Reproducible artifact references across environments
\end{itemize}

All tools share common CLI conventions (\texttt{--seed} for reproducibility, \texttt{--intelligence} for model selection, \texttt{--format} for output control) and integrate through Flyte for LLM operations, ensuring consistent behavior across the toolchain.

\section{Usage Examples}

\subsection{Create and Compile an ETL Data Processing Workflow}

\begin{verbatim}
dagify compose --type plain_text -f yaml --ascii \
  "Fetch data from API → clean data → load into PostgreSQL" \
  > pipeline_text.yaml

dagify compile pipeline_text.yaml -t pure --ascii --intelligence 10 \
   > pipeline_form.yaml

# Step 2: Execute the compiled workflow with dagent
dagent execute pipeline_form.yaml --tools=python,psql

\end{verbatim}

\begin{figure}[htbp]
  \centering
  \begin{subfigure}[b]{0.48\textwidth}
    \centering
    \includegraphics[width=\linewidth]{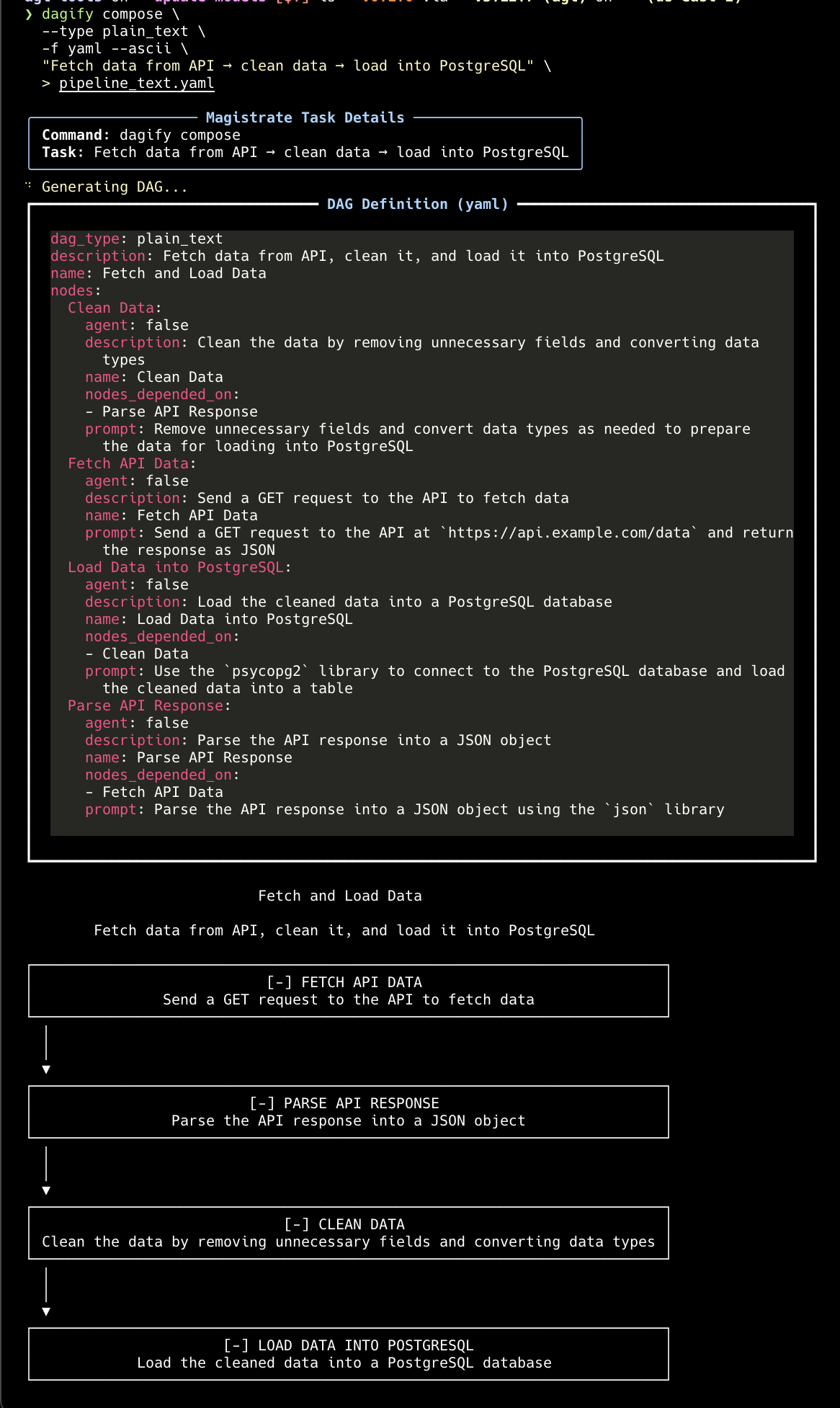}
    \caption{Output of \texttt{dagify compose} showing the generated DAG structure.}
    \label{fig:dagify-compose-output}
  \end{subfigure}
  \hfill
  \begin{subfigure}[b]{0.48\textwidth}
    \centering
    \includegraphics[width=\linewidth]{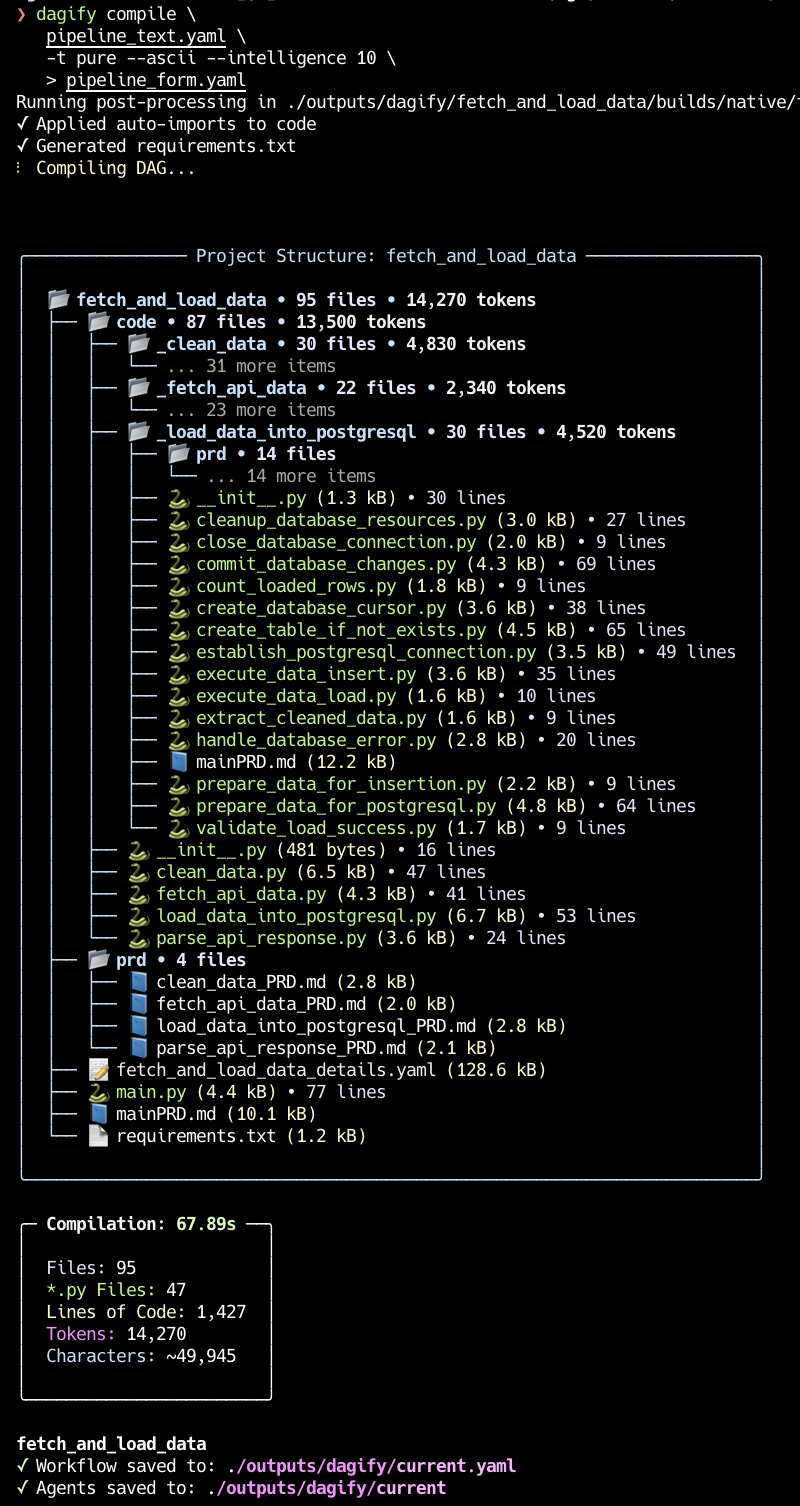}
    \caption{Output of \texttt{dagify compile} showing the code DAG file structure.}
    \label{fig:dagify-compile-output}
  \end{subfigure}
  \caption{Comparison of outputs from \texttt{dagify compose} and \texttt{dagify compile}.}
  \label{fig:dagify-compose-vs-compile}
\end{figure}

\subsection{Schema and Data Management}

\begin{verbatim}
# Generate a db schema for a blog with posts and comments
schemagin compose --format=json -f dbml \
    "Blog posts & comments feel

# Apply schema to database and generate test data
agiwrite schema --target-dbfer=duckdb://orders.db < schema.yaml
datagin synthesize "1000 orders" schema.yaml --rows=1000 \
  --output-agilink agilink://orders

# Transform data with anonymization
datagin transform "Remove PII" agilink://orders agilink://orders_clean
\end{verbatim}

\subsection{End-to-End Example: Analytics Pipeline}


\begin{verbatim}
# 1. Design schema
schemagin compose "Sales analytics schema" -f yaml > sales.yaml

# 2. Create workflow
dagify compose -t spec "ingest CSV -> validate -> aggregate -> report" \
  > pipeline.yaml

# 3. Resolve and compile
dagify resolve pipeline.yaml --type-floor=code > pipeline.code.yaml
dagify compile pipeline.code.yaml --build-target=wdl > pipeline.wdl

# 4. Execute with monitoring
dagent validate pipeline.wdl
dagent execute pipeline.wdl --jit=dynamic --tools=python,duckdb
\end{verbatim}



\begin{figure}[htbp]
  \centering
   \begin{subfigure}[b]{0.48\textwidth}
     \centering
     \includegraphics[width=\linewidth]{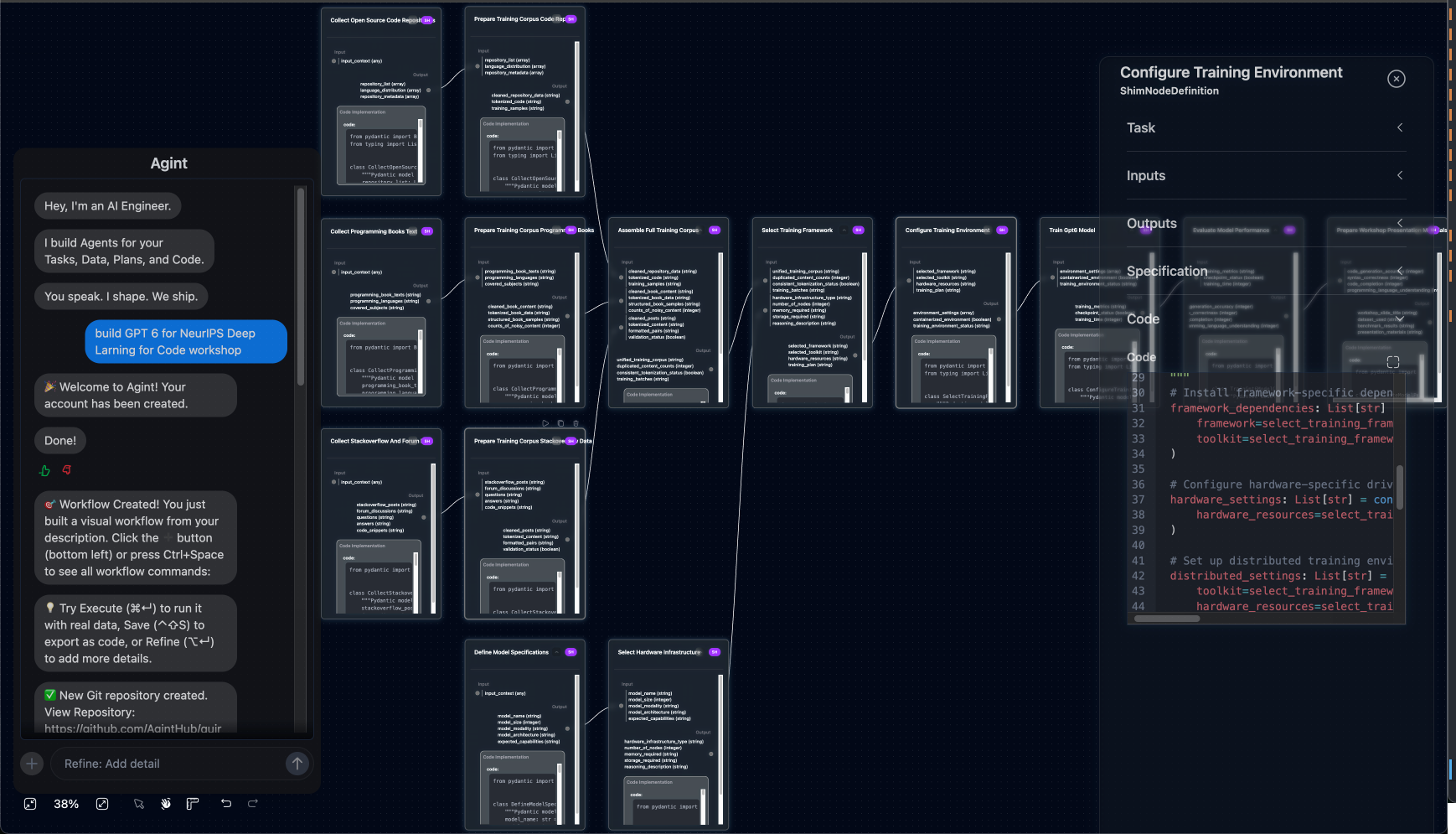}
     \caption{Agint Flow GUI.}
     \label{fig:delightful-coast}
   \end{subfigure}
  \hfill
  \begin{subfigure}[b]{0.48\textwidth}
    \centering
    \includegraphics[width=\linewidth]{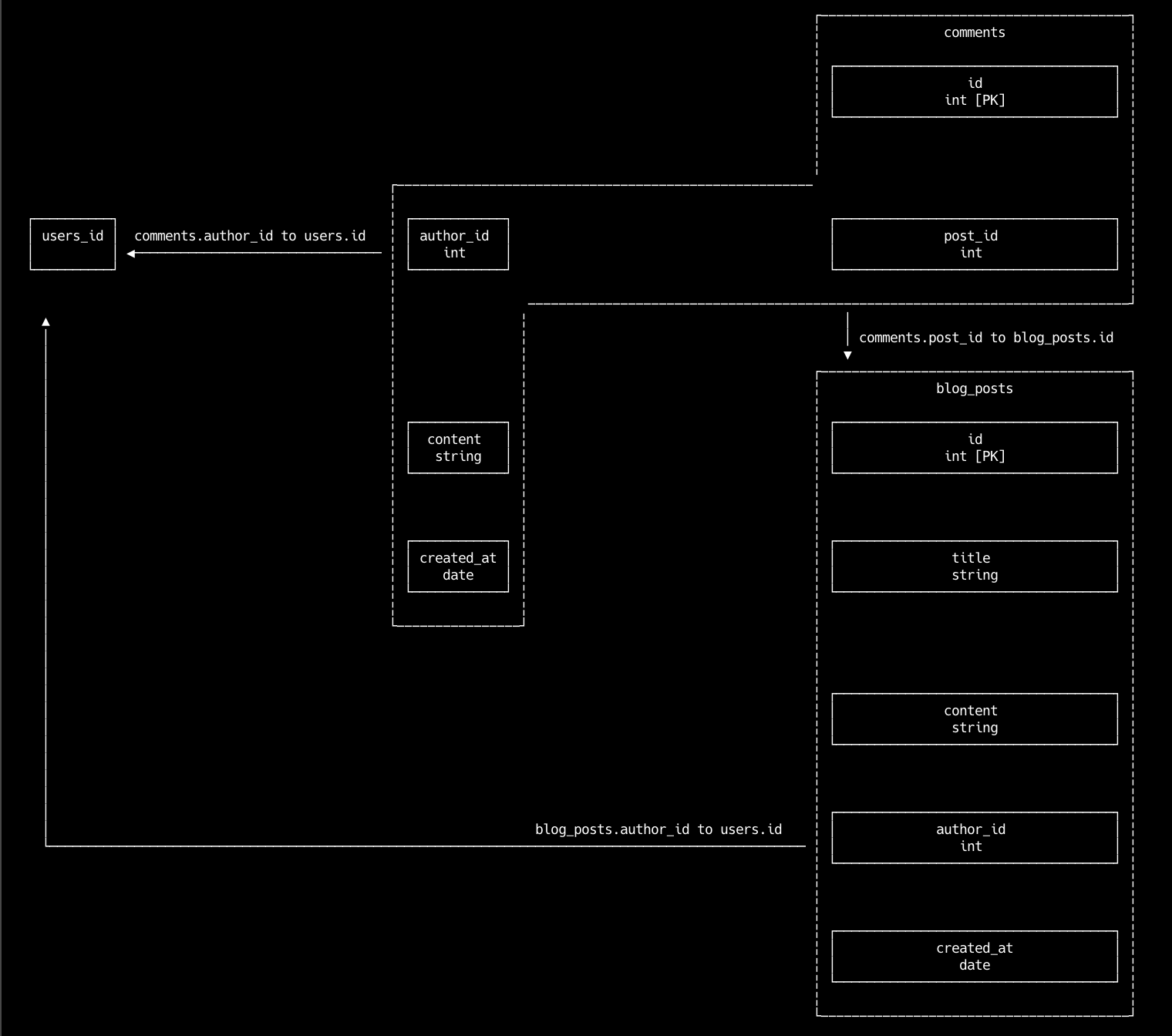}
    \caption{Screenshot of Automated Schema Generation via Agint CLI}
    \label{fig:daily-scripting}
  \end{subfigure}
  \label{fig:coast-vs-scripting}
\end{figure}

\section{Limitations and Societal Impacts}

\subsection{Limitations}

While Agint demonstrates significant advances in graph-based code generation, several limitations warrant discussion:

\textbf{Model Dependency}: The system's effectiveness depends heavily on the quality and availability of underlying LLMs. Performance degrades when models lack domain-specific knowledge or when rate limits restrict concurrent execution. The multi-provider routing in Flyte mitigates but does not eliminate this dependency.

\textbf{Scalability Constraints}: Although the DAG structure enables parallel execution, very large workflows (thousands of nodes) may encounter memory limitations during compilation and execution. The current implementation has been tested primarily on workflows with hundreds of nodes.

\textbf{Type System Limitations}: The \texttt{PrimitiveType} system constrains data to basic types (\texttt{str}, \texttt{int}, \texttt{float}, \texttt{bool}) and their lists. Complex data structures require serialization, potentially limiting expressiveness for certain domains.

\textbf{Evaluation Gaps}: While we demonstrate the system's capabilities through examples and case studies, comprehensive evaluation against established benchmarks remains future work. Existing benchmarks like SWE-bench~\citep{jimenez2023swebench} for repository-level issue resolution, ML-Bench~\citep{tang2023mlbench} for machine learning workflows, and Commit0~\citep{zhao2025commit0} for library generation would provide valuable comparative assessment of Agint's graph-based approach versus traditional methods.

\subsection{Societal Impacts}

Agint could democratize programming by enabling domain experts to specify complex workflows in natural language, potentially accelerating software development cycles through parallel generation and typed guarantees. The effect-aware execution model promotes reproducible AI-assisted development. However, concerns include potential displacement of entry-level programming jobs and organizational dependency on LLM providers.

\section{Conclusion and Future Work}

\subsection{Conclusion}

We presented Agint, an agentic graph compiler that transforms natural language specifications into executable code through hierarchical type floor-driven generation and hybrid execution. By treating code generation as a graph compilation problem, Agint addresses fundamental limitations of current LLM-based coding assistants: context management, reliability, and composability.

Our key contributions include:
\begin{itemize}
\item A six-tier type floor system (\texttt{TEXT} $\to$ \texttt{TYPED} $\to$ \texttt{SPEC} $\to$ \texttt{STUB} $\to$ \texttt{SHIM} $\to$ \texttt{PURE}) that enables incremental refinement and execution at any resolution level
\item Locality-preserving DAG compilation that enables parallel generation and reduces context requirements
\item A hybrid runtime supporting prefine, dynamic, and predict execution modes with effect-aware rollback
\item A composable Unix-style toolchain (dagify, dagent, schemagin, datagin) unified through the \texttt{agilink://} addressing system
\end{itemize}

The system demonstrates that compiler techniques—type systems, intermediate representations, and optimization passes—can effectively structure AI-powered code generation, making it more reliable and scalable than monolithic approaches.

\subsection{Future Work}

\textbf{Extended Type Systems}: Future versions will support richer type systems including algebraic data types, generics, and effect types to enable more precise specifications and better integration with strongly-typed languages.

\textbf{Formal Verification Integration}: Formal verification tools at the \texttt{SPEC} floor may provide stronger correctness guarantees. We envision automatic generation of proof obligations from specifications.

\textbf{Learning from Execution}: Feedback loops where execution results will improve future compilations may create self-improving systems. Failed executions would generate training data for model fine-tuning.

\textbf{Benchmark Suite}: We will evaluate Agint on established benchmarks including SWE-bench~\citep{jimenez2023swebench} ~\citep{yang2024swebenchmultimodal}, ML-Bench~\citep{tang2023mlbench}, and Commit0~\citep{zhao2025commit0} to quantify the benefits of the graph-based compilation approach over monolithic generation methods.

Agint represents a step toward treating AI code generation as a structured compilation problem rather than a text generation task. By grounding natural language in typed graphs and providing incremental refinement mechanisms, we enable a human-AI collaboration at scale in software development. As demand and utilization of LLMs increase, improve, the principles demonstrated here—locality, typing, and effect awareness—will become increasingly important for building reliable, scalable software engineering agents.







\begin{ack}
  We thank the reviewers and organizers of the DL4C Workshop at NeurIPS 2025 for their constructive feedback and valuable discussions that helped improve this work.

  We would also like to sincerely thank \textbf{Ahmed El-Kishky} for his encouragement and for believing in us throughout this project.

  This work was supported solely by internal research efforts at Agint and received no external financial support.

\end{ack}

\bibliographystyle{plain}
\bibliography{references}

@article{arora2024masai,
  title={MASAI: Modular Architecture for Software-Engineering AI Agents},
  author={Arora, Daman and others},
  year={2024},
  note={Available at \href{https://arxiv.org/abs/2406.11638}{arXiv:2406.11638}}
}

@article{besta2024graph,
  title={Graph of Thoughts: Solving Elaborate Problems with Large Language Models},
  author={Besta, Maciej and others},
  year={2024},
  note={Available at \href{https://arxiv.org/abs/2305.16582}{arXiv:2305.16582}}
}

@article{chen2021evaluating,
  title={Evaluating Large Language Models Trained on Code},
  author={Chen, Mark and others},
  year={2021},
  note={Available at \href{https://arxiv.org/abs/2107.03374}{arXiv:2107.03374}}
}

@article{chen2024funcoder,
  title={Divide-and-Conquer Meets Consensus: Unleashing the Power of Functions in Code Generation},
  author={Chen, Jingchang and others},
  year={2024},
  note={Available at \href{https://arxiv.org/abs/2405.20092}{arXiv:2405.20092}}
}

@article{gu2025aiforse,
  title={Challenges and Paths Towards AI for Software Engineering},
  author={Gu, Alex and others},
  year={2025},
  note={Position paper, available at \href{https://arxiv.org/abs/2503.22625}{arXiv:2503.22625}}
}

@article{hong2023metagpt,
  title={MetaGPT: Meta Programming for a Multi-Agent Collaborative Framework},
  author={Hong, Shuyue and others},
  year={2023},
  note={Available at \href{https://arxiv.org/abs/2308.00352}{arXiv:2308.00352}}
}

@article{hu2025qualityflow,
  title={QualityFlow: An Agentic Workflow for Program Synthesis Controlled by LLM Quality Checks},
  author={Hu, Yaojie and others},
  year={2025},
  note={Available at \href{https://arxiv.org/abs/2501.17167}{arXiv:2501.17167}}
}

@article{key2022trustworthy,
  title={Toward Trustworthy Neural Program Synthesis},
  author={Key, Darren and others},
  year={2022},
  note={Available at \href{https://arxiv.org/abs/2210.00848}{arXiv:2210.00848}}
}

@article{le2024codechain,
  title={CodeChain: Towards Modular Code Generation Through Chain of Self-Revisions with Representative Sub-Modules},
  author={Le, Hung and others},
  year={2024},
  note={Available at \href{https://arxiv.org/abs/2310.08992}{arXiv:2310.08992}}
}

@article{li2022alphacode,
  title={Competition-Level Code Generation with AlphaCode},
  author={Li, Yujia and others},
  year={2022},
  note={Available at \href{https://arxiv.org/abs/2203.07814}{arXiv:2203.07814}}
}

@article{li2024autokaggle,
  title={AutoKaggle: A Multi-Agent Framework for Autonomous Data Science Competitions},
  author={Li, Ziming and others},
  year={2024},
  note={Available at \href{https://arxiv.org/abs/2410.20424}{arXiv:2410.20424}}
}

@article{liu2024agents4se,
  title={Large Language Model-Based Agents for Software Engineering: A Survey},
  author={Liu, Junwei and others},
  year={2024},
  note={Available at \href{https://arxiv.org/abs/2409.02977}{arXiv:2409.02977}}
}

@article{muendler2025typeconstrained,
  title={Type-Constrained Code Generation with Language Models},
  author={Mündler, Niels and others},
  year={2025},
  note={Workshop paper, DL4C @ ICLR 2025, available at \href{https://arxiv.org/abs/2504.09246}{arXiv:2504.09246}}
}

@article{qian2024chatdev,
  title={ChatDev: Communicative Agents for Software Development},
  author={Qian, Chen and others},
  year={2024},
  note={Available at \href{https://arxiv.org/abs/2307.07924}{arXiv:2307.07924}}
}

@article{wang2024agents,
  title={Agents in Software Engineering: Survey, Landscape, and Vision},
  author={Wang, Yanlin and others},
  year={2024},
  note={Available at \href{https://arxiv.org/abs/2409.09030}{arXiv:2409.09030}}
}

@article{yao2023treeofthought,
  title={Tree of Thoughts: Deliberate Problem Solving with Large Language Models},
  author={Yao, Shunyu and others},
  year={2023},
  note={Available at \href{https://arxiv.org/abs/2305.10601}{arXiv:2305.10601}}
}

@article{zenkner2025taskdecomposition,
  title={Shedding Light on Task Decomposition in Program Synthesis: The Driving Force of the Synthesizer Model},
  author={Zenkner, Janis and others},
  year={2025},
  note={Accepted at ICLR 2025 Workshop, available at \href{https://arxiv.org/abs/2503.08738}{arXiv:2503.08738}}
}

@article{zhang2024dei,
  title={Diversity Empowers Intelligence: Integrating Expertise of Software Engineering Agents},
  author={Zhang, Kexun and others},
  year={2024},
  note={Available at \href{https://arxiv.org/abs/2408.07060}{arXiv:2408.07060}}
}

@article{jimenez2023swebench,
  title={SWE-bench: Can Language Models Resolve Real-World GitHub Issues?},
  author={Jimenez, Carlos E. and others},
  year={2023},
  note={Available at \href{https://arxiv.org/abs/2310.06770}{arXiv:2310.06770}}
}

@article{yang2024swebenchmultimodal,
  title={SWE-bench Multimodal: Do AI Systems Generalize to Visual Software Domains?},
  author={Yang, John and others},
  year={2024},
  note={Available at \href{https://arxiv.org/abs/2410.03859}{arXiv:2410.03859}}
}

@article{tang2023mlbench,
  title={ML-Bench: Evaluating Large Language Models and Agents for Machine Learning Tasks on Repository-Level Code},
  author={Tang, Xiangru and others},
  year={2023},
  note={Available at \href{https://arxiv.org/abs/2311.09835}{arXiv:2311.09835}}
}

@inproceedings{zhao2025commit0,
  title={Commit0: Library Generation from Scratch},
  author={Zhao, Wenting and others},
  booktitle={The Thirteenth International Conference on Learning Representations},
  year={2025},
  note={Available at \href{https://openreview.net/forum?id=MMwaQEVsAg}{OpenReview}}
}



\section{Technical Appendices and Supplementary Material}

\subsection{Interactive Demo and Documentation}

We provide comprehensive supplementary materials at the following locations:

\begin{itemize}
\item \textbf{Interactive Web Demo}: \url{https://flow.AgintAI.com} - A web-based interface demonstrating Agint Flow GUI with visual DAG editing, workflow execution, and real-time compilation from natural language specifications.

\item \textbf{API Documentation}: \url{https://api.AgintAI.com} - Complete API reference for programmatic access to Agint's compilation and execution services, including REST endpoints, authentication, and usage examples.
\end{itemize}

\subsection{Example Generated Code}

An example of Agint-generated workflow code for training a large language model workflow can be found at:
\begin{itemize}
\item \url{https://github.com/AgintHub/nifty-wilson/blob/agint/outputs/dagify/train_large_language_model/main.py}
\end{itemize}

This demonstrates the compiled output from natural language specification to executable Python code, showcasing the type floor progression and modular decomposition of complex workflows into manageable components.


\newpage
\section*{NeurIPS Paper Checklist}

\begin{enumerate}

\item {\bf Claims}
    \item[] Question: Do the main claims made in the abstract and introduction accurately reflect the paper's contributions and scope?
    \item[] Answer: \answerYes{}
    \item[] Justification: The abstract and introduction accurately present Agint as an agentic graph compiler that transforms natural language into executable code through type floors (TEXT→TYPED→SPEC→STUB→SHIM→PURE), hybrid execution modes (prefine/dynamic/predict), and a Unix-style toolchain. All claims are substantiated in the technical sections.

\item {\bf Limitations}
    \item[] Question: Does the paper discuss the limitations of the work performed by the authors?
    \item[] Answer: \answerYes{}
    \item[] Justification: Section 6.1 explicitly discusses limitations including model dependency (performance depends on LLM quality/availability), scalability constraints (tested primarily on hundreds of nodes, not thousands), and type system limitations (constrained to primitive types and their lists).

\item {\bf Theory assumptions and proofs}
    \item[] Question: For each theoretical result, does the paper provide the full set of assumptions and a complete (and correct) proof?
    \item[] Answer: \answerNA{}
    \item[] Justification: This is a systems paper presenting a software architecture and implementation, not a theoretical paper with formal proofs.

    \item {\bf Experimental result reproducibility}
    \item[] Question: Does the paper fully disclose all the information needed to reproduce the main experimental results of the paper to the extent that it affects the main claims and/or conclusions of the paper (regardless of whether the code and data are provided or not)?
    \item[] Answer: \answerYes{}
    \item[] Justification: The paper provides detailed system architecture, type floor definitions, execution modes, and concrete CLI examples showing how to use each component (dagify, dagent, schemagin, datagin). While no quantitative experiments are presented, the system design and workflow examples are described with sufficient detail for implementation.

\item {\bf Open access to data and code}
    \item[] Question: Does the paper provide open access to the data and code, with sufficient instructions to faithfully reproduce the main experimental results, as described in supplemental material?
    \item[] Answer: \answerNo{}
    \item[] Justification: The paper describes a system architecture and implementation approach but does not provide open-source code. The supplementary material at flow.AgintAI.com provides additional documentation and demos rather than reproducible code.

\item {\bf Experimental setting/details}
    \item[] Question: Does the paper specify all the training and test details (e.g., data splits, hyperparameters, how they were chosen, type of optimizer, etc.) necessary to understand the results?
    \item[] Answer: \answerNA{}
    \item[] Justification: The paper does not include experiments requiring training or testing. It presents a system architecture and compilation framework.

\item {\bf Experiment statistical significance}
    \item[] Question: Does the paper report error bars suitably and correctly defined or other appropriate information about the statistical significance of the experiments?
    \item[] Answer: \answerNA{}
    \item[] Justification: The paper does not include experiments with statistical results. It focuses on system design and architecture.

\item {\bf Experiments compute resources}
    \item[] Question: For each experiment, does the paper provide sufficient information on the computer resources (type of compute workers, memory, time of execution) needed to reproduce the experiments?
    \item[] Answer: \answerNA{}
    \item[] Justification: The paper does not include computational experiments. It describes system architecture and tools.
    
\item {\bf Code of ethics}
    \item[] Question: Does the research conducted in the paper conform, in every respect, with the NeurIPS Code of Ethics \url{https://neurips.cc/public/EthicsGuidelines}?
    \item[] Answer: \answerYes{}
    \item[] Justification: The research presents a software system for code generation without involving human subjects, sensitive data, or ethically concerning applications.

\item {\bf Broader impacts}
    \item[] Question: Does the paper discuss both potential positive societal impacts and negative societal impacts of the work performed?
    \item[] Answer: \answerYes{}
    \item[] Justification: Section 6.2 (Societal Impacts) discusses both positive impacts (democratizing programming for domain experts, accelerating development cycles) and negative impacts (potential displacement of entry-level programming jobs, organizational dependency on LLM providers).
    
\item {\bf Safeguards}
    \item[] Question: Does the paper describe safeguards that have been put in place for responsible release of data or models that have a high risk for misuse (e.g., pretrained language models, image generators, or scraped datasets)?
    \item[] Answer: \answerNA{}
    \item[] Justification: The paper does not release pretrained models, image generators, or scraped datasets. While the system includes effect-aware execution and rollback capabilities for safe code execution, these are operational features rather than release safeguards for high-risk assets.

\item {\bf Licenses for existing assets}
    \item[] Question: Are the creators or original owners of assets (e.g., code, data, models), used in the paper, properly credited and are the license and terms of use explicitly mentioned and properly respected?
    \item[] Answer: \answerYes{}
    \item[] Justification: The system uses commercial LLM APIs. No external code or datasets requiring special licensing are incorporated.

\item {\bf New assets}
    \item[] Question: Are new assets introduced in the paper well documented and is the documentation provided alongside the assets?
    \item[] Answer: \answerNo{}
    \item[] Justification: The paper introduces a new system (Agint) with novel components as a demo (Flyte, Dagify, Dagent, etc.), it does not release these as downloadable assets. The paper provides architectural documentation and usage examples, but the actual software implementation is not made publicly available. The paper is a Technical and Demo paper.

\item {\bf Crowdsourcing and research with human subjects}
    \item[] Question: For crowdsourcing experiments and research with human subjects, does the paper include the full text of instructions given to participants and screenshots, if applicable, as well as details about compensation (if any)? 
    \item[] Answer: \answerNA{}
    \item[] Justification: The research does not involve crowdsourcing or human subjects. It presents a software system architecture.

\item {\bf Institutional review board (IRB) approvals or equivalent for research with human subjects}
    \item[] Question: Does the paper describe potential risks incurred by study participants, whether such risks were disclosed to the subjects, and whether Institutional Review Board (IRB) approvals (or an equivalent approval/review based on the requirements of your country or institution) were obtained?
    \item[] Answer: \answerNA{}
    \item[] Justification: The research does not involve human subjects and therefore does not require IRB approval.

\item {\bf Declaration of LLM usage}
    \item[] Question: Does the paper describe the usage of LLMs if it is an important, original, or non-standard component of the core methods in this research? Note that if the LLM is used only for writing, editing, or formatting purposes and does not impact the core methodology, scientific rigorousness, or originality of the research, declaration is not required.
    \item[] Answer: \answerYes{}
    \item[] Justification: The paper extensively describes LLM usage as a core component. Section 2.4 details how Flyte orchestrates multiple LLM providers with automatic failover and hierarchical structured generation through Hydantic. LLMs are fundamental to transforming natural language to code across all type floors.

\end{enumerate}

\end{document}